\def\ARXIVVERSION{1}
\documentclass[sigconf,nonacm]{acmart}
\long\def\ARXIVAUTHORS{%
  \author{Jie Li}
  \orcid{0000-0001-6898-5086}
  \affiliation{%
    \institution{Independent Researcher}
    \city{Melbourne}
    \country{Australia}
  }
  \email{hey.jieli@gmail.com}

  \author{Dudu Luo}
  \orcid{0000-0002-0547-7516}
  \affiliation{%
    \department{School of Business, Law, Humanities and Pathways}
    \institution{University of Southern Queensland}
    \country{Australia}
  }
  \email{dudu.luo@unisq.edu.au}

  \author{Jiayang Niu}
  \orcid{0009-0003-7432-8156}
  \affiliation{%
    \department{School of Computing Technologies}
    \institution{RMIT University}
    \city{Melbourne}
    \country{Australia}
  }
  \email{s4068570@student.rmit.edu.au}

  \author{Ke Deng}
  \orcid{0000-0002-1008-2498}
  \affiliation{%
    \department{School of Computing Technologies}
    \institution{RMIT University}
    \city{Melbourne}
    \country{Australia}
  }
  \email{ke.deng@rmit.edu.au}

  \author{Yongli Ren}
  \orcid{0000-0002-3137-9653}
  \affiliation{%
    \department{School of Computing Technologies}
    \institution{RMIT University}
    \city{Melbourne}
    \country{Australia}
  }
  \email{yongli.ren@rmit.edu.au}
  \renewcommand{\shortauthors}{Li et al.}
}
\ifdefined\ARXIVVERSION
\else
\documentclass[sigconf,review,anonymous]{acmart}
\fi

\usepackage{amsmath,booktabs,multirow}
\usepackage{graphicx}
\usepackage{xcolor}
\usepackage{enumitem}

\ifdefined\ARXIVVERSION
\setcopyright{none}
\renewcommand{\shortauthors}{Li et al.}
\else
\renewcommand{\shortauthors}{Anonymous Author(s)}
\fi
\newcommand{\system}{\textsc{FunnelAudit}}

\begin{document}

\title{FunnelAudit: Responsibility Auditing in Multi-Route Recommender Systems}
\ifdefined\ARXIVVERSION
\ARXIVAUTHORS
\else
\author{Anonymous Author(s)}
\affiliation{\institution{Anonymous Institution}}
\fi

\begin{abstract}
  Modern recommender systems assemble a displayed list through multiple retrieval
  routes followed by allocation, fusion, and ranking. This makes a particular
  inclusion or exclusion difficult to audit. Route overlap can hide a control's
  effect from a one-at-a-time ablation, while failing to recompute downstream
  stages evaluates a counterfactual that the serving system would never execute.

  We introduce \system{}, an executable accountability framework for these
  incidents. An accountability contract registers the disputed Top-$K$ event,
  the controls and owners in scope, their permitted reference actions, and the
  stages that must be replayed. \system{} constructs the outcome of every
  permitted control configuration and applies graded actual responsibility to
  find the smallest factual-preserving contingency that makes each control
  pivotal. The resulting certificate records the contingency and both serving
  executions needed to check the judgment. We instantiate the framework in
  two-stage, nine-route funnels with fixed union, weighted quota allocation, or
  weighted reciprocal-rank fusion, followed by SASRec ranking.

  Across 258,809 user--target incidents from three real interaction datasets,
  4.24--16.24\% admit at least one responsible control under the registered
  contracts. Among responsible incident--control pairs, 92.55--99.64\% require a
  nonempty contingency, and leave-one-control-out recovers only 0.36--7.45\%.
  Moreover, serving policies whose factual outcomes differ on only 0.31--2.39\%
  of incidents produce 21.44--54.05\% Jaccard distance between responsible-route
  sets on matched exclusions. Independent literal replay reproduces 9,121,792
  target--world outcomes exactly; exhaustive scan and a generic mixed-integer
  linear program agree with every sampled judgment. These results show that
  recommender accountability requires explicit serving and replay semantics
  together with checkable incident-level evidence, not another aggregate
  importance score.
\end{abstract}

\ccsdesc[500]{Information systems~Recommender systems}
\ccsdesc[300]{Social and professional topics~Accountability}
\ccsdesc[300]{Computing methodologies~Causal reasoning and diagnostics}
\keywords{recommender systems, accountability, actual responsibility,
multi-route retrieval, reciprocal rank fusion, causal auditing}

\maketitle

\section{Introduction}

A recommender turns a large catalog into a personalized Top-$K$ list. Under
online latency constraints, large systems commonly narrow the catalog through
a serial \emph{multi-stage} cascade and use parallel \emph{multi-route}
generators within retrieval~\cite{covington2016youtube,wang2020cold}. The routes
cover complementary interests; allocation and ordering rules decide which
candidates survive~\cite{huang2025multichannel}. The displayed list therefore
reflects several interacting components and policies rather than the output of
one model.

This architecture creates a concrete accountability problem: why did a target
item appear in, or fail to appear in, a particular list? The displayed list
shows what happened, but not which route or policy made the difference. Because
routes overlap and later stages respond to changes upstream, the answer cannot
be read from a final score or a simple provenance record.

\begin{figure*}[t]
  \centering
  \includegraphics[width=\textwidth]{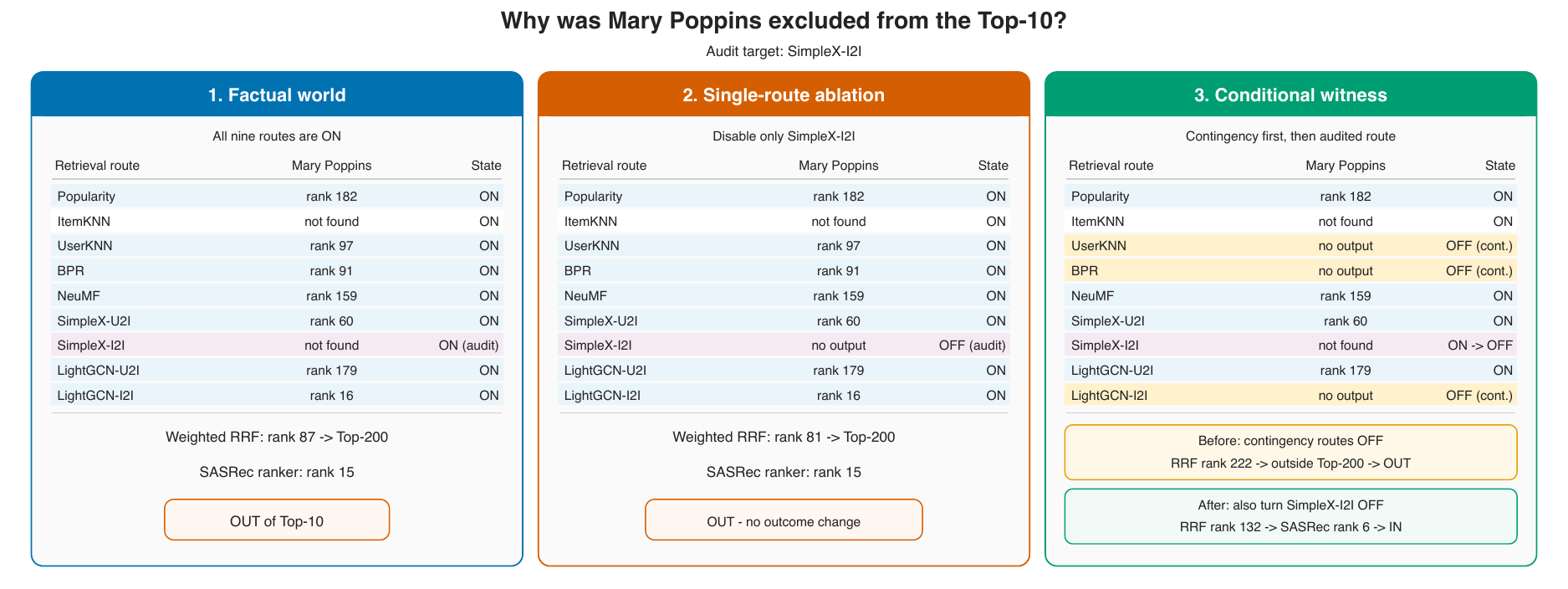}
  \caption{Three views of the same MovieLens-1M exclusion audit. The factual
    world shows the observed serving path. The single-route ablation disables
    SimpleX-I2I alone and leaves the target excluded. The conditional witness
    pairs a contingency-only world with a world that additionally disables
    SimpleX-I2I; only this final comparison changes the target from exclusion to
  inclusion.}
  \Description{Three side-by-side panels list all nine retrieval routes and
    whether Mary Poppins appears in each Top-200 list. The factual panel keeps all
    routes on and ends at SASRec rank 15. The single-route-ablation panel disables SimpleX-I2I and
    also ends at rank 15. The conditional-witness panel marks UserKNN, BPR, and
    LightGCN-I2I as contingency routes and SimpleX-I2I as the audited route. With
    only the contingency disabled, Mary Poppins has RRF rank 222 and is outside
    the candidate set. Disabling SimpleX-I2I as well changes the RRF rank to 132
  and the SASRec rank to 6, placing the movie in the final Top-10.}
  \label{fig:audit-incident}
\end{figure*}

Figure~\ref{fig:audit-incident} illustrates the problem in a nine-route system.
Each retrieval route returns up to 200 movies for one user. Reciprocal-rank
fusion (RRF) merges the overlapping lists and retains 200 candidates; SASRec
then ranks those candidates and displays the first 10
~\cite{cormack2009rrf,kang2018sasrec}. Route provenance is insufficient here.
Several routes may nominate the target, while a route that never nominates it
can still support competitors that occupy the limited candidate slots.
Disabling a route also changes the input seen by fusion and ranking. The audit
must therefore follow the complete serving path rather than inspect one route
or the final score in isolation.

We instantiate the example with the pipeline used in our experiments and a real
MovieLens-1M interaction history~\cite{harper2015movielens}. MovieLens-1M
supplies the user--item history; our experimental pipeline produces the route
outputs. For the user in Figure~\ref{fig:audit-incident}, \emph{Mary Poppins
(1964)} does not enter the final Top-10. We call an absent target an
\emph{exclusion} and a displayed target an \emph{inclusion}. Seven routes
retrieve the movie, but the ranker places it 15th. Suppose we audit the
SimpleX-I2I route. It does not retrieve Mary Poppins, and disabling it alone
still leaves the movie outside the displayed list. This one-route check finds
no effect.

That check asks only whether the audited route is decisive in the original
configuration. A responsibility audit also asks whether it becomes decisive
after the smallest set of other route changes that, by themselves, leave Mary
Poppins excluded. For this user, that set disables UserKNN, BPR, and
LightGCN-I2I. These three changes form the \emph{contingency}. From exactly that
configuration, we then disable SimpleX-I2I. The route never retrieved Mary
Poppins, but it had supplied RRF support to competing movies. Removing that
support lets Mary Poppins cross the fusion cutoff, after which SASRec ranks it
sixth. The complete pipeline runs immediately before and after this final
change form the \emph{witness}: the first preserves the exclusion, while the
second changes it to an inclusion.

This comparison depends on what happens after a route is switched off.
Recomputing RRF and ranking produces the witness above; retaining their previous
outputs would ask a different counterfactual question. An audit must therefore
state the disputed event, the controls and owners in scope, the permitted change
for each control, and the later stages that respond. Following the actor--forum
account~\cite{bovens2007accountability}, we record these choices in an
\emph{accountability contract}. Each registered control is tied to an owner,
while the reviewing forum defines the question and permitted interventions.
The audit changes only those control states and recomputes the pipeline from the
first affected stage onward. We call this \emph{stage-faithful replay}. Compared
configurations use the same trained models and policy definitions; only the
permitted control states and their downstream consequences may differ. The
returned certificate identifies the contingency and witness so that another
implementation can check the technical judgment. It does not infer intent,
liability, or a remedy.

Figure~\ref{fig:audit-incident} uses rank-aware fusion to expose one form of
route interaction. Our study also covers weighted quota allocation. A quota
allocator is a candidate-construction component that determines how many
top-ranked items each active route may contribute. In that serving policy, we
audit its weighted-quota-versus-bypass state alongside route availability. The
three serving policies and their audit scopes are specified in
Section~\ref{sec:serving-model} before the common accountability model.

We use graded actual responsibility as the judgment rule. A control is
responsible if it becomes decisive after an outcome-preserving contingency.
Its degree of responsibility is higher when fewer additional control changes
are needed~\cite{chockler2004responsibility}. This theory assumes that the
outcome of every allowed intervention has already been defined. A multi-stage
recommender does not provide those counterfactual outcomes directly: switching
off a route can redistribute quota, alter the fused candidate set, and change
the ranking of the target and its competitors. Our main formulation
contribution is to define a complete Top-$K$ result for each permitted control
configuration. The contract specifies those configurations, and stage-faithful
replay constructs their outcomes. Together they turn the general causal rule
into an executable item-level audit.

\system{} executes this contract in two steps. It first uses stage-faithful replay
to record the Top-$K$ outcome under every permitted control configuration, then
searches those outcomes for each control's smallest valid contingency. The
search directly implements graded actual responsibility rather than introducing
a new generic causality algorithm. FunnelAudit's methodological contribution is
the recommender-specific construction of the counterfactual worlds, controls,
and replayable evidence to which that existing rule is applied. In our nine-
and ten-control settings, an independently implemented serving replay, per-pair
exhaustive search, and a mixed-integer linear program (MILP) remain feasible
~\cite{nemhauser1988integer}. We use serving replay to validate world
construction and the other two methods to validate responsibility extraction.
We do not claim production-scale enumeration of all configurations.

We instantiate the audit in a two-stage, nine-route pipeline reconstructed from
a public multi-channel design
~\cite{huang2025multichannel}. The experiments cover full-route union, weighted
quota allocation, and weighted RRF candidate fusion,
each followed by the same SASRec ranker~\cite{kang2018sasrec}. We attribute
responsibility to route availability and, only under weighted quota allocation,
the allocator's apply-versus-bypass control. RRF and SASRec are
recomputed after an upstream change but are not themselves audited. Auditing
fusion or ranker controls remains outside our empirical claim.

Therefore, our contributions are:

\begin{itemize}[leftmargin=*]
  \item \textbf{A recommender-specific accountability formulation.} We define
    the incident, controls, admissible interventions, stage-faithful replay, and
    certificate required to apply graded actual responsibility to interacting
    route-availability and allocator-policy controls.

  \item \textbf{An executable audit framework.} We develop \system{}, which
    evaluates all configurations permitted by the contract and extracts exact
    judgments and certificates. Independent serving replay checks world
    construction, while per-pair scan and MILP check responsibility extraction.

  \item \textbf{A real-data measurement study.} We reconstruct a documented
    nine-route suite on MovieLens-1M, Amazon Beauty, and UCI Online Retail II.
    Across three serving policies and 258,809 user--target incidents,
    92.55--99.64\% of responsible pairs require a nonempty contingency, while
    single-control ablation recovers only 0.36--7.45\% of responsible
    incident--control pairs. Allocation and fusion semantics materially change
    the judgments.
\end{itemize}

\section{Background and Related Work}

\subsection{Accountability, audits, and evidence}

Accountability is a relationship, not a property of an explanation alone. In
the actor--forum account, an actor must explain and justify conduct to a forum
that can question and judge it~\cite{bovens2007accountability}. The relationship
therefore needs an object, a standard, and a forum with defined authority
~\cite{wieringa2020account}. This matters for machine-learning systems because
one outcome may involve many people and components~\cite{cooper2022algorithmic}.
In our setting, the actor is the owner of a registered serving control, while
the forum fixes the incident, permitted interventions, and required evidence.
This places technical responsibility within a review process without treating
causal responsibility as moral or legal blame.

Prior work supplies important parts of this process. Transparency and
interrogation expose how an algorithm operates~\cite{diakopoulos2016accountability};
internal auditing, reviewability, traceability, and decision provenance organize
records across a system lifecycle
~\cite{raji2020closing,cobbe2021reviewable,kroll2021traceability,singh2019decision}.
Cryptographic systems can even prove that an agreed audit function was executed
over committed data and models without revealing them
~\cite{waiwitlikhit2024trustless}. These works strengthen the integrity and
governance of an audit, but take its technical question as given. They do not
specify which serving controls may be changed, which downstream stages must
respond, or what constitutes responsibility for one disputed recommendation.

Recommender-system audits operate at several other units of analysis.
Bellog\'{i}n and Said address the accountability of experimental research
through reproducible data processing and evaluation
~\cite{bellogin2021accountability}. Risk-scenario audits examine platform-level
harms and mitigation over time~\cite{messmer2023auditing}. Sharma et al. define
causal metrics for reachability, stability, and user agency in multi-step
recommendation dynamics~\cite{sharma2024unified}; RecAudit tests whether a
sequential recommender retains a user's behavioral profile after data
revocation~\cite{zhu2026forget}. Li et al. instead ask which users or items
affect a system-level fairness measure~\cite{li2025acfr}. These are substantive
audit questions, but their objects are research practice, platform risk, user
agency, privacy, or aggregate fairness. Our unit of judgment is different: one
factual Top-$K$ inclusion or exclusion and the registered serving controls that
may have produced it. The gap is therefore in the causal object and permitted
interventions, not simply in the choice of metric.

\subsection{Actual responsibility and recommendation explanations}

Chockler and Halpern grade an actual cause by the smallest contingency needed
to make it critical~\cite{chockler2004responsibility}. Database causality
adapts this idea to query answers and non-answers and studies the complexity of
responsibility~\cite{meliou2010complexity,meliou2010functional}. These works
supply the pivotality and contingency standard once the causal variables and
outcome function have been fixed. Our task is the recommender-specific
instantiation: define versioned serving controls and reference actions, replay
each admissible joint action through the affected stages, and return a
checkable certificate for a Top-$K$ event.

Recommendation explanations usually intervene on data presented to, or used
by, the model. Halpern--Pearl causality has been adapted to rated items,
stakeholder profiles, and aggregate recommendations
~\cite{verdeaux2020causality}. PRINCE finds minimal user actions whose removal
changes a recommendation~\cite{ghazimatin2020prince}; CountER alters item
aspects~\cite{tan2021counter}; and learned counterfactual explainers remove
subsets of user history~\cite{yao2022counterfactual}. More recent work studies
targeted, rank-aware edits to a user's interaction sequence
~\cite{kim2026trace}. These methods answer what input would have produced a
different recommendation. Li et al.'s ACFR similarly changes interaction data,
but its explanandum is an aggregate fairness measure rather than one serving
incident~\cite{li2025acfr}.

Why-not explanations are the closest neighbors to our motivating question.
For collaborative filtering, Stratigi et al. diagnose a missing item through
data, score, and model parameters~\cite{stratigi2020whynot}. Attolou et al.
promote a missing graph recommendation by adding or removing user-rooted edges
~\cite{attolou2024whynot}. Top-$K$ evaluation of counterfactual explainers has
also received separate study~\cite{mohammadi2025beyond}. Our target event can
likewise be an absent item, but our candidate causes are not user actions, graph
edges, or unconstrained model parameters. They are owned serving controls with
registered reference actions. Their interventions can change
candidate membership and competitor support, so the question cannot be answered
from the final model score alone. The output is an owner-linked
actual-responsibility certificate for the observed event, rather than an input
edit intended to obtain a preferred list.

\subsection{Multi-stage serving, fusion, and contribution scores}

Multi-stage serving first reduces a large catalog and then applies more
expensive ordering models. Candidate generation followed by ranking is a
standard industrial pattern~\cite{covington2016youtube}; pre-ranking and
reranking can add further stages~\cite{wang2020cold}. Multi-route retrieval is
parallel rather than serial: several generators cover complementary parts of
the catalog before their outputs are combined. Recent work on compositional
recommender fairness explicitly models interactions between retrieval, scoring,
and serving components, but evaluates and optimizes system-level utility and
equity rather than responsibility for a realized item-level event
~\cite{hsu2025models}. Huang et al. optimize fusion and route budgets in a
multi-channel retriever~\cite{huang2025multichannel}. We reconstruct their nine
route interfaces as a serving benchmark; their objective is retrieval quality,
not an audit judgment over the controls that produced one inclusion or
exclusion.

Reciprocal rank fusion (RRF) combines ranked lists through reciprocal positions
without calibrating their original score scales~\cite{cormack2009rrf}. When an
active route changes, its nominations and rank contributions change, so the
fused candidate set must be recomputed. We treat this response as part of the
serving replay rather than claiming RRF itself as an audited control or a
contribution of this paper.

Shapley and Banzhaf values summarize average marginal contribution
across coalitions~\cite{lundberg2017shap,wang2023banzhaf}. They are useful
credit-allocation baselines, but they answer a different question from minimum
factual contingency. A route may have high recall, learned weight, or average
coalitional contribution and still not be pivotal for the incident under
review. Conversely, a route hidden by redundancy can have a small average
contribution yet possess a short factual-preserving witness. FunnelAudit
therefore reports cause status, the minimum contingency, and a checkable
witness rather than treating a contribution score as accountability.

Together, these strands leave one operational gap: none links a registered
Top-$K$ incident to owned serving-control interventions, replays every affected
stage, and returns a minimum factual-preserving witness. FunnelAudit fills that
gap. It does not claim a new accountability theory, fusion rule, or general
explanation method.

\section{Serving Model and Audit Scenarios}
\label{sec:serving-model}

Modern recommender systems are often multi-stage: retrieval narrows a large
catalog before more expensive ordering stages process the surviving items. We
study a two-stage instance. Stage 1 runs several retrieval routes and constructs
a candidate set. Each active route $r$ returns an ordered list $P_r(L)$ of at
most $L$ items. Stage 2 applies a downstream ranker to that candidate set and
returns the displayed Top-$K$. When an intervention changes Stage 1, the audit
recomputes every affected Stage-1 operation and reruns Stage 2 on the resulting
candidate set.

Within this two-stage pipeline, we study three retrieval-stage serving
scenarios: fixed union (C1), weighted quota allocation (C2), and weighted RRF
candidate fusion (C3). Route availability is registered as a control in all
three. C2 additionally registers an allocator-policy control, denoted $A$. Its
factual action applies the registered quota rule; its reference action bypasses
that rule and uses fixed union. The allocator determines how much of each active
route's ranked output may enter the candidate set; it is neither a route nor the
final ranker. RRF and the downstream ranker are replayed when their inputs
change, but are not themselves audited controls. Our empirical scope is
therefore responsibility for retrieval routes and, in C2, the quota allocator.
The nine route interfaces follow the public multi-channel design of Huang et
al.~\cite{huang2025multichannel}. We use the same route outputs and downstream
ranker across scenarios so that the candidate-construction policy, rather than
a different set of retrievers, determines how their serving worlds differ.
Section~\ref{sec:experimental-setup} gives the datasets, learned weights, route
depth, candidate budget, and ranker used in the empirical instantiation.

\subsection{C1: Fixed union}

Every active route contributes its complete prefix $P_r(L)$. The pipeline unions
and deduplicates those items before final ranking. Removing a route can change
candidate membership, so the union and ranker are reconstructed after every
route intervention. C1 contains no quota allocator or RRF step. It isolates
redundancy among route outputs.

\subsection{C2: Weighted quota allocation}

Let $S$ be the set of active routes and let $w_r$ be the registered weight of
route $r$. In the factual state of $A$, the allocator renormalizes the weights of
the active routes and assigns the integer prefix quota

\begin{equation}
  q_r(S)=\left\lfloor B\frac{w_r}{\sum_{j\in S}w_j}+\frac{1}{2}\right\rfloor.
  \label{eq:quota}
\end{equation}

If $S=\varnothing$, the allocator returns an empty candidate set. Otherwise,
the selected prefixes are unioned and deduplicated without refill. Consequently,
disabling a route changes not only its own output but also the quotas of the
surviving routes. In the reference state of $A$, candidate construction bypasses
the quota rule and uses the C1 fixed union. Any resulting change in candidate-set
size is part of that registered intervention.

\subsection{C3: Weighted RRF candidate fusion}

Reciprocal rank fusion combines ranked lists through rank positions rather than
their raw, potentially incomparable scores~\cite{cormack2009rrf}. Every active
route contributes its complete prefix $P_r(L)$. For item $i$, the retrieval-side
fusion score under active-route set $S$ is

\begin{equation}
  s_{\mathrm{RRF}}(i;S)=\sum_{r\in S:\,i\in P_r(L)}
  \frac{w_r}{b_{\mathrm{RRF}}+\operatorname{rank}_r(i)}.
  \label{eq:rrf}
\end{equation}

The pipeline recomputes this score over all unique nominations and retains the
best $B$ items for the downstream ranker. Disabling a route can therefore change
target support, competitor support, fused candidate membership, and final order.
C3 has no allocator control: RRF is the fixed candidate-construction rule and is
replayed after route interventions. Figure~\ref{fig:audit-incident} instantiates
this scenario.

\section{Accountability Contract and Formal Model}
\label{sec:formal-model}

\subsection{Accountability contract}

With the serving scenarios fixed, we can now state the common audit question.
Before responsibility can be computed, the audit must specify the
counterfactual it intends to evaluate. In Figure~\ref{fig:audit-incident}, for example,
the factual event is that a particular movie is absent from one user's Top-10.
Whether a route is responsible depends on which controls may be changed, what
action represents changing each control, and which downstream stages are rerun.
Two audits that make different choices here can reach different judgments even
when they start from the same observed ranking. We call this explicit
specification the \emph{accountability contract}.

For one request and one item under review, we write the contract as

\begin{equation}
  \mathcal{A}_{x,t}=(\mathcal{F},x,t,K,y_0,\mathcal{C},\omega,\alpha,\sigma,
  \operatorname{Serve},\tau),
  \label{eq:audit-instance}
\end{equation}

The first five entries identify who asks the question and which factual event
is at issue. $\mathcal{F}$ is the reviewing forum, $x$ is the recorded request,
$t$ is the item under review, $K$ is the display cutoff, and
$y_0\in\{0,1\}$ records whether $t$ appeared in the factual Top-$K$ list. The
remaining entries define the counterfactual scope. $\mathcal{C}$ lists the
serving controls that may be questioned; $\omega(c)$ records the owner of
control $c$; and $\alpha(c)=(a_c^0,a_c^{\mathrm{ref}})$ records its factual
action and the one alternative action permitted by the forum. The stage map
$\sigma(c)$ says where replay starts, while the versioned
$\operatorname{Serve}$ function reconstructs the final list. Finally, $\tau$
is the technical judgment rule. Throughout this paper, $\tau$ is the
minimum-contingency pivotality rule defined in
Section~\ref{sec:technical-standard}. In our experiments, a reference action
disables a route or bypasses the allocator. Changing the controls, reference
actions, replay rule, or judgment rule changes the audit question itself,
rather than giving another solver for the same question.

The completed audit returns one record for every registered control:

\begin{equation}
  \operatorname{Audit}(\mathcal{A}_{x,t})=
  \{(c,a_c^0,\omega(c),J_c,\rho_c,E_c):c\in\mathcal{C}\},
  \label{eq:audit-output}
\end{equation}

Here $J_c$ says whether the factual action $c=a_c^0$ is a technical cause of
the registered Top-$K$ event. The degree $\rho_c$ is larger when fewer other
permitted interventions are needed to make that action decisive, and $E_c$ is
the replayable evidence for the judgment. The next two subsections define these
worlds and judgments precisely. Thus the output makes a conditional statement:
under this contract, the factual action of control $c$ was or was not
technically responsible for this item-level event. It does not determine
intent, blame, liability, or remedy. Any organizational response remains with the reviewing forum; \system{}
supplies only the registered technical judgment and its supporting evidence.

\subsection{Registered controls and counterfactual serving worlds}

For request $x$, let $\mathcal{C}=\{c_1,\ldots,c_m\}$ be the registered binary
serving controls. Each control has a stable identifier, an owner, a factual
action, one permitted reference action, and an earliest affected stage. The
causal query concerns the factual control--action pair $(c,a_c^0)$; we say that
``control $c$ is responsible'' only as shorthand for responsibility of this
factual action. A route control, for example, is factually available and becomes
unavailable under its reference action. In C2, the allocator-policy control
$A\in\mathcal{C}$ either applies the registered weighted-quota policy or
bypasses that policy. C1 and C3 do not register $A$.

A counterfactual world is indexed by the set $D\subseteq\mathcal{C}$ of
controls placed in their reference state. Its joint action profile is

\begin{equation}
  a_D(c)=
  \begin{cases}
    a_c^{\mathrm{ref}}, & c\in D,\\
    a_c^0, & c\notin D.
  \end{cases}
  \label{eq:action-profile}
\end{equation}

The factual world is $D=\varnothing$. In our experiments, $D$ is implemented as
a reference-state bit mask: a route bit disables that route, whereas the $A$
bit selects the registered allocator bypass. Each binary control therefore
represents one declared factual-versus-reference comparison. Auditing a second
alternative value of a multi-valued policy requires a separate contract.

The current extractor uses the full Boolean domain, so every
$D\subseteq\mathcal{C}$ must describe an executable joint action. If two
registered actions cannot coexist, the forum must restrict or revise the
registry before applying this extractor. Handling constrained or non-Boolean
intervention domains is outside the scope of this paper.

\paragraph{Stage-faithful replay.}
Changing a control may invalidate outputs produced later in the pipeline. We
therefore start at the earliest stage affected by any intervention in $D$ and
recompute every downstream component whose input or governing policy may have
changed. For example, disabling a route changes candidate construction, so the
applicable allocator or fusion step and the final ranker are recomputed.
Bypassing the allocator changes its candidate set, so the ranker is rerun on
that set. A component keeps the same trained parameters and policy definition,
but its output is recomputed when its input changes. Outputs that provably
cannot depend on the intervention may be reused.

This replay produces a deterministic ranked list and a binary target outcome:

\begin{equation}
  L_x(D)=\operatorname{Serve}(x;a_D),\qquad
  Y_{x,t}(D)=\mathbf{1}\{t\in\operatorname{TopK}_K(L_x(D))\}.
  \label{eq:serving-outcome}
\end{equation}

For the present implementation, a contract is well formed only when every
registered joint action is executable, replay is deterministic and
stage-faithful, and the reconstructed factual world satisfies
$y_0=Y_{x,t}(\varnothing)$. The candidate-construction, deduplication,
allocation, fusion, ranking, and tie-breaking operations present in the
registered pipeline are applied exactly as specified. The audit therefore
assigns responsibility relative to this finite intervention model, not to
every possible cause of the event in the real world.

\paragraph{Stage scope.}
The contract can register a pre-ranker, ranker, reranker, or business rule when
the forum supplies a stable reference action and enough state to replay its
downstream stages. Our experiments register only route availability and, in C2,
the allocator's weighted-quota-versus-bypass action. RRF and SASRec respond to
these interventions but are not themselves assigned responsibility.

\subsection{Technical standard and judgment}
\label{sec:technical-standard}

We specialize the criticality principle of graded actual
responsibility~\cite{chockler2004responsibility} to the finite control registry
above. We do not posit a complete structural causal model of the recommender
and its environment. Instead, the contract fixes the causal variables, their
one permitted reference action, and the resulting serving outcome in every
admissible Boolean world. Responsibility is therefore evaluated only over this
registered intervention domain.

For the factual action of control $c$ to be responsible, it must become
decisive after changing some other controls. Those other changes form a
contingency $W$. They must leave the registered event unchanged on their own;
placing $c$ in its reference state as well must flip the event. Formally, let

\begin{equation}
  \mathcal{W}_c=\left\{W\subseteq\mathcal{C}\setminus\{c\}:\;
    Y_{x,t}(W)=y_0,\;
  Y_{x,t}(W\cup\{c\})\neq y_0\right\}.
  \label{eq:witness}
\end{equation}

Each $W\in\mathcal{W}_c$ is a valid contingency. The ordered pair of replayed
worlds $(W,W\cup\{c\})$ is its witness: the first preserves the factual event,
and the second shows that changing $c$ makes the difference. We define

\begin{align}
  \kappa(c;x,t)&=
  \begin{cases}
    \min_{W\in\mathcal{W}_c}|W|, & \mathcal{W}_c\neq\varnothing,\\
    \infty, & \mathcal{W}_c=\varnothing,
  \end{cases}\\
  J_c&=\mathbf{1}[\mathcal{W}_c\neq\varnothing],\\
  \rho(c;x,t)&=
  \begin{cases}
    1/(1+\kappa(c;x,t)), & \text{if }\kappa(c;x,t)<\infty,\\
    0, & \text{otherwise.}
  \end{cases}
\end{align}

We write $\rho_c=\rho(c;x,t)$. Each query concerns one factual control--action
pair. The controls in $W$ identify the circumstances under which that action
becomes pivotal; they are not folded into the queried cause. Minimizing $|W|$
gives the graded judgment: a factual action that needs fewer supporting changes
receives a larger $\rho$.

The same definition applies to factual inclusion and exclusion. When
$\kappa=0$, changing the audited control alone flips the event; this is the
ordinary but-for test. When $\kappa>0$, changing that control alone does not
flip the event, but it becomes pivotal after an outcome-preserving contingency
changes other controls.
Several minimum contingencies may be valid. The registry's stable control-ID
order assigns each control a fixed bit position; among equal-size
contingencies, we choose the smallest integer mask as the contingency reported
in the canonical certificate. This tie-break makes the evidence deterministic
without asserting that the minimum witness is unique. It does not affect
$J_c$, $\kappa$, or $\rho$.

\paragraph{Shared responsibility.}
When the allocator control $A$ is registered alongside the routes, all of them
belong to the same counterfactual domain. A route's minimum contingency may
contain $A$, and $A$'s factual action may become pivotal after one or more routes
are disabled. An audit that separates the two layers rules out these
contingencies by design and therefore answers a different question. Here,
``shared'' refers to the joint route--allocator audit space, not to a group-cause
score: $J_c$ and $\rho_c$ remain per-control judgments. In serving
configurations without a registered allocator control, the union or fusion
mechanism is replayed but is not itself a cause variable.

\paragraph{Leave-one-control-out baseline.}
Leave-one-control-out (LOCO) applies the reference action to exactly one
registered control while all other controls remain in their factual states;
equivalently, it tests only the empty contingency $W=\varnothing$. In C1 and
C3, a LOCO test disables one route and replays the downstream pipeline. In C2,
it either disables one route and recomputes quotas over the remaining active
routes, or bypasses the allocator while leaving every route active. Changing a
route and the allocator together is therefore a nonempty-contingency test, not
LOCO.

\paragraph{Certificate.}
For $J_c=1$, the certificate identifies the incident, control action,
contingency $W$, the action profiles for $W$ and $W\cup\{c\}$, their target
ranks and outcomes, and the resulting $\kappa$ and $\rho$. Input hashes and
links to the factual and witness replays make these fields checkable. For
$J_c=0$, it instead identifies the complete outcome table and exhaustive edge
scan establishing that no valid contingency exists. Thus $(J_c,\rho_c)$ is the
technical judgment and $E_c$ is its supporting evidence.

\section{FunnelAudit: From Contract to Certificates}

Section~\ref{sec:formal-model} defines the audit question; \system{} computes
its answer in two steps. A scenario-specific procedure replays every permitted
control configuration and records whether the target appears in the Top-$K$.
An extractor then finds a minimum contingency and assembles the corresponding
judgment and certificate. The extractor is not a new generic causality
algorithm. It is a finite-domain implementation of graded actual responsibility
~\cite{chockler2004responsibility}, following the contingency-set search view
used in database causality~\cite{meliou2010complexity}. We use this method
because the contract gives each registered control one binary reference action,
so the permitted worlds form a Boolean lattice and every witness is an edge in
that lattice.

\subsection{Replaying the contract as an outcome lattice}

\begin{figure}[t]
  \centering
  \includegraphics[width=\columnwidth]{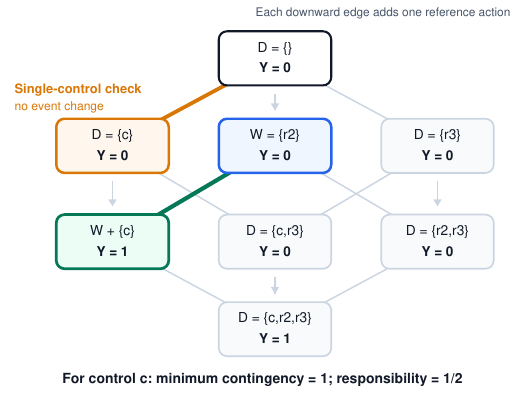}
  \caption{A Boolean outcome lattice for three binary controls. Here $c$ is the
    audited control, while $r_2$ and $r_3$ are the other registered controls.
    Orange shows that changing $c$ alone leaves the event at $Y=0$.
    The blue node $W=\{r_2\}$ is a one-control contingency that also preserves
    the event. From that world, additionally changing $c$ reaches the green node
    and flips the event to $Y=1$. Thus $(W,W\cup\{c\})$ is a minimum witness,
    $\kappa_c=1$, and $\rho_c=1/2$. Every node is produced by stage-faithful
  replay of the registered serving contract.}
  \Description{An eight-node Boolean lattice for controls c, r2, and r3. The
    factual node has outcome zero. Disabling c alone also has outcome zero. A
    highlighted contingency node for r2 has outcome zero, while the adjacent
  node that additionally disables c has outcome one, forming a minimum witness.}
  \label{fig:lattice}
\end{figure}

For one request--target pair, the set $D\subseteq\mathcal{C}$ identifies the
controls placed in their reference state. The compiler runs the registered
pipeline for each $D$ and stores $Y_{x,t}[D]$. The resulting $2^m$-entry table
is a Boolean lattice: each node is one replayed world, nodes on the same level
change the same number of controls, and an edge joins worlds that differ in
exactly one control. Figure~\ref{fig:lattice} illustrates this construction for
three binary controls. The factual world is $D=\emptyset$, and moving down one
edge applies one additional reference action. Changing $c$ alone does not alter
the outcome, but $c$ becomes pivotal after the contingency $W=\{r_2\}$. This
highlighted edge is a concrete instance of the witness pair $W$ and
$W\cup\{c\}$ in Eq.~\ref{eq:witness}.

The three scenarios share this output format but require different compilers.
For C1, the compiler unions and deduplicates the active route prefixes and then
reruns the ranker. For C2, it first recomputes the quotas of the active routes,
then rebuilds and ranks the candidate set. For C3, it recomputes weighted RRF
over the complete active-route prefixes, retains the best 200 fused candidates,
and reruns the ranker. The shared output format therefore preserves the serving
differences in Section~\ref{sec:serving-model}.

Exact world construction is exponential in the number of registered controls.
C1 and C3 contain $2^9=512$ worlds per request; C2 contains
$2^{10}=1{,}024$. These sizes are feasible for the deliberately small binary
registries studied here. Larger or non-binary control domains remain outside
our empirical scope.

\subsection{Judgment and evidence extraction}

Once the lattice has been built, the responsibility definition no longer
depends on whether an outcome came from union, quota allocation, or RRF. For
control $c$, the extractor compares each $W$ not containing $c$ with the
adjacent world $W\cup\{c\}$. The pair is a witness if $W$ preserves the factual
event and adding $c$ flips it.

The extractor considers contingencies in increasing order of $|W|$. The first
valid one therefore gives the minimum $\kappa$ and hence the responsibility
degree $\rho=1/(1+\kappa)$. Equal-size contingencies are considered in
increasing integer-mask order, implementing the deterministic tie-break in
Section~\ref{sec:technical-standard}. If a valid pair is found, the extractor
returns $J_c=1$, its minimum size $\kappa_c$, the derived degree $\rho_c$, and
the two replay records required by the certificate. If none exists, it returns
$J_c=0$ together with the complete outcome table that supports the negative
judgment.

We use the lattice-edge scan as the primary extractor because it matches the
object produced by the compiler. Once the $2^m$ registered worlds have been
replayed, every possible witness is already represented by an adjacent lattice
edge. One traversal can retain the minimum witness and certificate for every
registered control in the incident. A per-pair exhaustive scan instead repeats
the contingency search for each control, while the generic selector MILP used
in this work must be instantiated for each incident--control pair after the same
outcomes have been compiled. Neither alternative avoids stage-faithful world
construction. We therefore use the direct edge scan as the transparent,
deterministic reference extractor and retain per-pair scan and MILP as
independent exact checks. This choice improves workload reuse, but it does not
change the exponential $2^m$ world-enumeration boundary.

\subsection{Independent evidence verification}

We separately test two failure modes: an incorrect search of the outcome table
and incorrect construction of the serving worlds.

\paragraph{Extractor checks.}
A per-pair exhaustive scan searches the outcome lattice through an independent
implementation. A generic selector MILP~\cite{nemhauser1988integer} instead
selects a factual-preserving pivotal edge of minimum size, using one binary
variable per candidate contingency. Both are exact where applied, but both
start from the compiled outcomes. Their agreement checks extraction, not world
construction.

\paragraph{Independent serving-world check.}
An independently implemented literal replay bypasses the outcome-lattice
construction code. Starting from versioned route outputs and model scores, it
reconstructs registered worlds, including quota renormalization, RRF Top-200
membership, and SASRec ordering, without importing the compiler. Agreement on
the resulting Top-$K$ outcomes checks serving semantics and world construction;
the extractor checks above test minimum-witness search separately. RQ5 reports
the sampling and coverage of these checks.

\section{Experimental Setup}
\label{sec:experimental-setup}

\subsection{Experimental accountability instantiation}

Public interaction data identify neither an organizational forum nor the owners
of serving controls. We therefore instantiate $\mathcal{F}$ as a fixed technical
review protocol and assign each route, and the C2 allocator control $A$, a
stable ID and owner role. Before measurement, the protocol fixes the incident,
reference actions, replay semantics, responsibility standard, certificate
schema, and independent checks. We evaluate the technical judgment and evidence
defined by the contract, without attributing outcomes to real employees or
studying deliberation, sanctions, or remedies.

\subsection{Research questions}

\begin{description}[leftmargin=0pt,labelindent=0pt]
  \item[RQ1: Judgment materiality.] Does the contract yield nontrivial
    responsibility judgments, and how much pivotality is contingency-hidden?
  \item[RQ2: Search completeness.] How much exact responsibility is missed when
    witness search is restricted to single-control, bounded, or randomly sampled
    contingencies?
  \item[RQ3: Serving semantics.] How do allocation and retrieval-side fusion
    change the responsible set and the form of a valid witness?
  \item[RQ4: Estimand specificity.] Can average-contribution scores substitute
    for the minimum-contingency judgment?
  \item[RQ5: Verification.] Does independent serving replay reproduce the
    compiled outcomes, and do per-pair scan and MILP reproduce the judgments?
\end{description}

\subsection{Datasets, splits, and targets}

We use MovieLens-1M~\cite{harper2015movielens}, Amazon Beauty 5-core
~\cite{mcauley2015amazon}, and UCI Online Retail II
~\cite{chen2012onlineretail}. MovieLens treats ratings $\geq4$ as positives and
uses per-user chronological 5:2:3 splits. Beauty treats reviews as implicit
positives; Retail retains non-cancelled, positive-price purchases and
deduplicates customer--product pairs. All datasets require at least ten retained
interactions per user; Beauty and Retail use global chronological 5:2:3 splits.

Models and routes use only training and validation data. For each audit user,
the query history is their train-plus-validation history. Every positive test
item that is absent from this history and present in the warm catalog defines a
separate user--target incident. All of a user's targets are evaluated against
the same held-out request: we do not add earlier test interactions to the
history when auditing a later target. This design covers the complete eligible
test relevance set without leaking test behavior into the request.

Let $\mathcal{I}$ denote these incidents. We evaluate every registered control
for every incident: $9|\mathcal{I}|$ incident--control pairs in C1 and C3 and
$10|\mathcal{I}|$ in C2. This is not an audit of arbitrary catalog items; every
target is a recorded positive test interaction. Test data select none of the
models, weights, thresholds, serving policies, or audit rules.
The upper block of Table~\ref{tab:setup-summary} reports both the number of
distinct users and the larger incident population.

\begin{table}[t]
  \centering
  \caption{Audit populations and mean factual candidate-pool sizes. C3 pre/post denotes the pool before and after the RRF cutoff.}
  \label{tab:setup-summary}
  \small
  \setlength{\tabcolsep}{3pt}
  \begin{tabular}{@{}lrrrr@{}}
    \toprule
    \multicolumn{5}{@{}c}{  \textit{Audit populations}} \\
    Dataset & Users & Incidents & Items & Interactions \\
    \midrule
    ML-1M & 5,950 & 174,888 & 3,421 & 574,619 \\
    Beauty & 2,660 & 11,564 & 10,429 & 91,824 \\
    Online Retail & 2,488 & 72,357 & 4,204 & 477,410 \\
    \addlinespace
    \toprule
    \multicolumn{5}{@{}c}{\textit{Mean factual candidate-pool size}} \\
    Dataset & C1 & C2 & C3 pre & C3 post \\
    \midrule
    ML-1M & 599.45 & 153.54 & 599.45 & 200.00 \\
    Beauty & 1,063.71 & 167.25 & 1,063.71 & 200.00 \\
    Retail & 763.41 & 131.40 & 763.41 & 200.00 \\
    \bottomrule
  \end{tabular}
\end{table}

\subsection{Nine-route funnel and ranker}

Nine is a benchmark choice, not an estimated industry average. We adopt the
nine-channel configuration of Huang et al.~\cite{huang2025multichannel} because
it spans several retrieval mechanisms while permitting exact intervention
enumeration. The routes are Popularity; ItemKNN and UserKNN
~\cite{sarwar2001itembased,resnick1994grouplens}; BPR~\cite{rendle2009bpr};
NeuMF~\cite{he2017neumf}; and U2I and I2I interfaces for both
SimpleX~\cite{mao2021simplex} and LightGCN~\cite{he2020lightgcn}. U2I scores
catalog items directly, whereas I2I expands items in the user's history; we
register the two interfaces separately. This list is also the order used for
all route vectors.

Neural routes are selected on validation data and refitted on training plus
validation. Every route returns $L=200$ unique unseen items with deterministic
ties. C1 unions these prefixes without a post-union cutoff. C2 uses $B=200$ as
the nominal budget from which route quotas are computed; deduplication can leave
fewer than 200 candidates. C3 instead applies weighted RRF and retains exactly
the best $B=200$ candidates. Thus $L$ is the per-route depth in all cases,
whereas $B$ is used only by C2 and C3. Candidate budgets in the hundreds are consistent
with public descriptions of large-scale funnels
~\cite{covington2016youtube,linkedin2024semantic}. We audit final Top-10
membership and use Top-20 as a sensitivity.

Before auditing, we verify legal, unique, unseen route outputs and credible
retrieval and ranking behavior. Full-union Recall@200 is .778, .277, and .556
on MovieLens, Beauty, and Retail, versus .556, .107, and .311 for the best
single route. For the SASRec ranker, candidate-restricted/full-catalog ratios
for Recall@10 and NDCG@10 range from .988 to 1.013; only the ratio can exceed
one when filtering removes high-scoring irrelevant items. These checks were
fixed and independently replayed before accountability measurement.

C2 route weights are selected using validation-only Cross-Entropy Method (CEM)
optimization~\cite{rubinstein1999cem,huang2025multichannel}. The resulting
weights and coalition-dependent quotas are fixed before test auditing; exact
vectors and optimization traces are retained in the artifact.
C3 uses the same learned weights and fixes $b_{\mathrm{RRF}}=60$, following
the value selected in the original RRF study~\cite{cormack2009rrf}. This
constant smooths differences between route positions and is distinct from the
candidate budget $B$. Test targets and accountability results select none of
these choices.

Within each dataset, one SASRec model~\cite{kang2018sasrec} scores all unseen
items. Stage 2 restricts those scores to the candidates produced by C1, C2, or
C3 and returns the Top-10. An intervention rebuilds the candidate set and
reapplies the same ranker without retraining. Table~\ref{tab:setup-summary} confirms the
intended candidate construction: all factual C3 unions exceed 200 nominations
and are reduced to exactly 200 candidates, so the RRF cutoff is active in every
request.

\subsection{Baselines and serving-policy comparisons}

\paragraph{Witness-search baselines.}
LOCO tests only the empty contingency $W=\varnothing$ ($\kappa=0$) across the
registered controls. Bounded-$k$
search exhaustively checks $|W|\leq1$ or $|W|\leq2$.
Random Search (RS) uniformly samples 8, 32, or 128 contingency sets per
incident--control pair. Larger RS budgets retain all contingencies examined by
smaller budgets, and every budget also checks $W=\varnothing$. All use the same
stage-faithful lattice, so a found witness is valid but may not be minimum,
while an unvisited witness can be missed.

\paragraph{Contribution-value baselines.}
We define coalition utility as preservation of the registered factual event,
$u(D)=\mathbf{1}\{Y_{x,t}(D)=y_0\}$, and compute absolute Shapley--Shubik and
Banzhaf values from the complete outcome lattice
~\cite{lundberg2017shap,wang2023banzhaf}. They use the same counterfactual
worlds as \system{}, but aggregate marginal changes over coalitions rather than
minimize an outcome-preserving contingency.

\paragraph{Serving-policy comparisons.}
We compare responsible route sets across C1, C2, and C3 and separately measure
the C2 route witnesses whose minimum contingency contains $A$. For C3, the
certificate in Figure~\ref{fig:audit-incident} also records whether the target
crosses the RRF candidate boundary on the pivotal edge. Outcome-specific
cross-case comparisons retain incidents included or excluded under both
policies and report factual-outcome disagreements separately. Each scenario is
a different registered serving policy, so these comparisons measure audit
sensitivity to serving semantics rather than the isolated causal effect of
turning an allocator or fusion method on.

\subsection{Reported quantities and evaluation criteria}

For an incident $(u,t)$ and control $c$, a \emph{responsible pair} has
$J_{u,t,c}=1$ if some possibly empty contingency $W$ preserves the factual
outcome but disabling $c$ as well changes it. The minimum $|W|$ is
$\kappa_{u,t,c}$: $\kappa=0$ is visible to LOCO, whereas $\kappa\geq1$ is
\emph{contingency-hidden}. Table~\ref{tab:reported-quantities} summarizes the
reported quantities and their denominators. An incident has only one factual
outcome, but a user with several test items may contribute both inclusion and
exclusion incidents.

\begin{table}[t]
  \centering
  \caption{Reported accountability quantities.}
  \label{tab:reported-quantities}
  \footnotesize
  \setlength{\tabcolsep}{3pt}
  \begin{tabular}{@{}p{.27\columnwidth}p{.67\columnwidth}@{}}
    \toprule
    Quantity & Definition and denominator \\
    \midrule
    Causal incidents & Incidents with at least one responsible control; micro rate divides by all audit incidents. \\
    Causal inclusions & Causal incidents with $y^0_{u,t}=1$; divided by all factual-inclusion incidents. \\
    Causal exclusions & Causal incidents with $y^0_{u,t}=0$; divided by all factual-exclusion incidents. \\
    Causal-user coverage & Users with at least one causal incident; divided by all audit users. \\
    User-macro rate & Each user's causal-incident fraction, averaged with equal weight across users. \\
    Responsible pairs & Incident--control pairs with $J_{u,t,c}=1$; mean $\rho$ and hidden share are computed over these pairs. \\
    Hidden share & Responsible pairs with $\kappa\geq1$, divided by all responsible pairs. \\
    \bottomrule
  \end{tabular}
\end{table}

To compare witness-search methods, \emph{responsible-pair recall} measures the
fraction of the exact responsible pairs that a method recovers. For
Shapley--Shubik and Banzhaf, we report Spearman correlation with $\rho$ over all
incident--control pairs and, over causal incidents, the expected probability
that a uniformly tie-broken maximum-score control is responsible. We report
both quantities separately for factual inclusion and exclusion: correlation
uses all pairs in the named stratum, whereas TopResp. uses only its causal
incidents. Verification uses separate pass/fail checks: literal replay requires
exact Top-$K$ outcome agreement, while extractor checks require exact cause
status, minimum contingency size, and canonical witness.

\section{Results}

\subsection{RQ1: Responsibility is common enough to matter and is usually hidden}

Table~\ref{tab:reported-quantities} defines the quantities used to answer RQ1;
Table~\ref{tab:main-new} reports the results. The unit of audit is one eligible
held-out user--item pair
$(u,t)$, not one user. Replaying the factual policy labels each pair an
\emph{inclusion incident} if $t$ appears in the final Top-10 and an
\emph{exclusion incident} otherwise. A user with several test items therefore
contributes several incidents. At this level, 4.24--16.24\% of incidents are
causal. Giving each user equal weight yields user-macro rates of
5.10--24.75\%, while 18.61--92.57\% of users have at least one causal
incident. These rates have different denominators and are not interchangeable.

The two factual outcomes behave differently. Every factual inclusion is causal
under the registered interventions. Starting from an included target, disabling
all routes eventually removes it; along that sequence, at least one route must
be the first control whose removal changes inclusion to exclusion. Hence every
included incident has at least one responsible control, although not every
route is responsible, and the causal-inclusion rate is 100\%.

Exclusions are asymmetric because the registered actions disable serving
controls rather than add a missing candidate or change the ranker. Removing a
route can sometimes promote an excluded target by removing competitors, but
only 2.44--10.20\% of factual exclusions admit such a witness. We call the
remainder \emph{contract-invariant exclusions}: the target stays outside the
Top-10 in every admissible control configuration. Accordingly,
89.80--97.56\% of factual exclusions remain unchanged under every intervention
permitted by their respective contracts and are certified non-attributable
within that scope. They are not missed by the extractor, nor do they imply that
no cause exists outside the registered controls and interventions.

At the control level, the audit identifies between 3,815 and 205,277
responsible incident--control pairs.
Among them, 92.55--99.64\% require a nonempty contingency and are therefore
hidden from a single-control ablation. Mean responsibility degree $\rho$
ranges from .153 to .272. RQ1 is therefore answered in both parts: the contract
produces a material, nontrivial set of judgments, and the large majority of
responsible pairs are contingency-hidden.

\begin{table}[t]
  \centering
  \caption{Exact responsibility at final $K=10$. ``Causal inc.'' is the micro
    incident rate; ``Macro'' averages each user's causal-incident fraction;
    ``Causal users'' is unique-user coverage; and inclusion and exclusion rates use
    the corresponding factual incidents as denominators. For count-based rows,
    cells report count (rate); ``Hidden'' is the percentage of responsible pairs
  with $\kappa\geq1$.}
  \label{tab:main-new}
  \scriptsize
  \begin{tabular*}{\columnwidth}{@{\extracolsep{\fill}}llrrr@{}}
    \toprule
    Dataset & Quantity & C1 & C2 & C3 \\
    \midrule
    \multirow{8}{*}{ML-1M}
    & Causal inc. & 20,800 (11.89\%) & 28,410 (16.24\%) & 22,563 (12.90\%) \\
    & Causal incl. & 11,483 (100\%) & 11,768 (100\%) & 11,786 (100\%) \\
    & Causal excl. & 9,317 (5.70\%) & 16,642 (10.20\%) & 10,777 (6.61\%) \\
    & Causal users & 5,274 (88.64\%) & 5,508 (92.57\%) & 5,321 (89.43\%) \\
    & User macro & 19.55\% & 24.75\% & 20.88\% \\
    & Resp. pairs & 129,414 & 205,277 & 146,590 \\
    & \textbf{Hidden} & \textbf{98.85\%} & \textbf{93.27\%} & \textbf{95.73\%} \\
    & Mean $\rho$ & .182 & .266 & .234 \\
    \midrule
    \multirow{8}{*}{Beauty}
    & Causal inc. & 665 (5.75\%) & 828 (7.16\%) & 786 (6.80\%) \\
    & Causal incl. & 224 (100\%) & 200 (100\%) & 205 (100\%) \\
    & Causal excl. & 441 (3.89\%) & 628 (5.53\%) & 581 (5.11\%) \\
    & Causal users & 495 (18.61\%) & 602 (22.63\%) & 568 (21.35\%) \\
    & User macro & 6.34\% & 8.02\% & 7.45\% \\
    & Resp. pairs & 3,815 & 6,109 & 4,934 \\
    & \textbf{Hidden} & \textbf{97.82\%} & \textbf{92.55\%} & \textbf{95.07\%} \\
    & Mean $\rho$ & .212 & .272 & .256 \\
    \midrule
    \multirow{8}{*}{Retail}
    & Causal inc. & 3,067 (4.24\%) & 3,653 (5.05\%) & 3,423 (4.73\%) \\
    & Causal incl. & 1,336 (100\%) & 1,480 (100\%) & 1,402 (100\%) \\
    & Causal excl. & 1,731 (2.44\%) & 2,173 (3.07\%) & 2,021 (2.85\%) \\
    & Causal users & 1,426 (57.32\%) & 1,514 (60.85\%) & 1,475 (59.28\%) \\
    & User macro & 5.10\% & 5.86\% & 5.53\% \\
    & Resp. pairs & 22,197 & 29,167 & 25,424 \\
    & \textbf{Hidden} & \textbf{99.64\%} & \textbf{95.65\%} & \textbf{96.83\%} \\
    & Mean $\rho$ & .153 & .231 & .222 \\
    \bottomrule
  \end{tabular*}
\end{table}

\subsection{RQ2: Local and budgeted searches are sound but incomplete}

Every baseline searches the same stage-faithful worlds, so every positive it
returns is a valid responsibility witness; the methods differ only in how many
exact responsible pairs their restricted searches recover.
Table~\ref{tab:search-new} reports this recall separately for factual inclusions
and exclusions. Unlike the incident rates in Table~\ref{tab:main-new}, each
denominator here contains only exact responsible incident--control pairs in the
corresponding outcome stratum.

LOCO recovers 0.39--18.84\% of responsible inclusion pairs and
0.34--5.19\% of responsible exclusion pairs. Exhaustively searching
contingencies of size at most two raises recall to 2.20--38.80\% and
4.20--23.09\%, respectively. RS-128 performs better, reaching
73.95--93.69\% for inclusions and
70.70--86.45\% for exclusions, but remains incomplete in every setting. The
relative difficulty of inclusion and exclusion varies with the serving policy
and dataset, so pooled recall can conceal meaningful differences. Pooling the
two strata recovers the aggregate LOCO range of 0.36--7.45\%. A failed
restricted search therefore does not establish that no responsible control
exists.

\begin{table}[t]
  \centering
  \caption{Recall (\%) of exact responsible pairs. Cells show
  inclusion/exclusion, separately normalized within each outcome stratum.}
  \label{tab:search-new}
  \scriptsize
  \resizebox{\columnwidth}{!}{%
    \begin{tabular}{@{}llcccccc@{}}
      \toprule
      Dataset & Case & LOCO & $\kappa\leq1$ & $\kappa\leq2$ & RS-8 & RS-32 & RS-128 \\
      \midrule
      \multirow{3}{*}{ML-1M} & C1 & 1.03/1.29 & 3.72/4.82 & 8.77/11.68 & 19.77/24.81 & 40.07/51.42 & 73.95/81.25 \\
      & C2 & 9.12/5.19 & 17.28/12.47 & 26.25/23.09 & 28.23/21.41 & 50.36/46.95 & 76.90/76.71 \\
      & C3 & 4.58/3.95 & 9.84/10.27 & 17.81/21.74 & 24.60/26.61 & 46.57/55.81 & 77.05/86.45 \\
      \midrule
      \multirow{3}{*}{Beauty} & C1 & 6.67/1.02 & 17.69/3.23 & 31.15/9.88 & 50.00/21.91 & 74.23/48.14 & 92.56/83.95 \\
      & C2 & 18.84/5.04 & 25.77/11.72 & 38.80/20.79 & 44.14/20.61 & 71.04/43.93 & 91.66/74.61 \\
      & C3 & 13.14/2.96 & 21.45/8.86 & 35.12/19.76 & 40.80/21.69 & 71.71/48.36 & 93.69/83.91 \\
      \midrule
      \multirow{3}{*}{Retail} & C1 & .39/.34 & .84/1.34 & 2.20/4.20 & 12.50/13.92 & 35.68/36.98 & 76.72/76.04 \\
      & C2 & 6.61/2.91 & 13.08/7.36 & 21.99/16.48 & 21.93/16.47 & 47.01/39.65 & 78.75/70.70 \\
      & C3 & 2.95/3.31 & 7.42/8.99 & 14.16/18.70 & 22.88/24.82 & 48.09/52.47 & 80.43/83.02 \\
      \bottomrule
    \end{tabular}%
  }
\end{table}
\subsection{RQ3: Allocation and fusion semantics change the audit}

C2 exposes cross-control responsibility. Table~\ref{tab:semantics-outcome}(a)
separates its two factual outcomes. The allocator is responsible in
18.94--35.00\% of inclusion incidents but only .92--3.48\% of exclusion
incidents. This difference does not make the allocator irrelevant to route
responsibility in exclusions: 10.55--14.37\% of responsible exclusion-route
pairs require the allocator in their minimum contingency. The corresponding
inclusion share is 5.43--12.24\%. A route-only audit cannot express either kind
of joint witness.

C3 changes a different part of the serving contract. Table~\ref{tab:setup-summary}
confirms that its factual RRF cutoff is active on every request.
Table~\ref{tab:semantics-outcome}(b) compares the exact route-responsibility
sets. For a meaningful outcome-specific comparison, Incl. and Excl.
retain only incidents with the same factual outcome under both cases; incidents
whose outcomes differ are reported separately. Factual disagreement is only
.31--2.39\%, yet matched-exclusion Jaccard distances reach 21.44--54.05\%.
Thus the serving policy can substantially change which routes are responsible
even when the target remains excluded under both policies. Matched-inclusion
distances are smaller (0--4.25\%) in these data. Allocation and fusion are
therefore not cosmetic labels for the same audit. Figure~\ref{fig:audit-incident}
shows one mechanism: the target remains nominated, but the pivotal edge moves
it across the RRF candidate boundary.

\begin{table}[t]
  \centering
  \caption{Outcome-stratified effects of allocation and fusion. In (a), cells
    show inclusion/exclusion counts (rates in \%); the two rates use factual incidents
    and responsible route pairs in their respective strata. In (b), Incl. and
    Excl. are Jaccard distances on incidents with matching outcomes under both
  cases; Outcome diff. is the factual-outcome disagreement rate.}
  \label{tab:semantics-outcome}
  \scriptsize
  \textit{(a) C2 cross-control effects}\par\vspace{1pt}
  \resizebox{\columnwidth}{!}{%
    \begin{tabular}{@{}lcc@{}}
      \toprule
      Dataset & Allocator-resp. incidents & Route pairs needing allocator \\
      \midrule
      ML-1M  & 2,229 (18.94) / 5,678 (3.48) & 8,451 (10.86) / 13,361 (11.17) \\
      Beauty & 70 (35.00) / 302 (2.66)       & 122 (12.24) / 681 (14.37) \\
      Retail & 311 (21.01) / 652 (.92)       & 603 (5.43) / 1,804 (10.55) \\
      \bottomrule
    \end{tabular}%
  }
  \vspace{3pt}

  \textit{(b) Route-responsibility Jaccard distance (\%)}\par\vspace{1pt}
  \resizebox{\columnwidth}{!}{%
    \begin{tabular}{@{}llrrrr@{}}
      \toprule
      Dataset & Cases & All & Incl. & Excl. & Outcome diff. (\%) \\
      \midrule
      \multirow{3}{*}{ML-1M}
      & C1--C2 & 37.61 & .00 & 54.05 & 2.39 \\
      & C1--C3 & 28.65 & .00 & 36.70 & 1.91 \\
      & C2--C3 & 31.13 & 2.20 & 44.24 & 1.43 \\
      \midrule
      \multirow{3}{*}{Beauty}
      & C1--C2 & 39.84 & .00 & 33.77 & 1.42 \\
      & C1--C3 & 41.56 & .00 & 37.49 & 1.17 \\
      & C2--C3 & 22.74 & 4.25 & 22.38 & .54 \\
      \midrule
      \multirow{3}{*}{Retail}
      & C1--C2 & 23.32 & .00 & 30.72 & .66 \\
      & C1--C3 & 20.10 & .00 & 26.39 & .31 \\
      & C2--C3 & 15.62 & .98 & 21.44 & .48 \\
      \bottomrule
    \end{tabular}%
  }
\end{table}

\subsection{RQ4: Average contribution does not replace incident responsibility}

RQ4 tests the most direct attribution alternative: can Shapley--Shubik or
Banzhaf scores stand in for FunnelAudit's responsibility judgment?
Table~\ref{tab:coalition-new} compares them using the same complete outcome
lattice as FunnelAudit, so every method sees the same counterfactual worlds.
Both baselines measure average coalition contribution. For a control $c$, they
aggregate the marginal change $u(D\cup\{c\})-u(D)$ over configurations $D$ of
the other controls. Shapley--Shubik weights each configuration by its
probability of preceding $c$ in a random intervention order; Banzhaf weights
all configurations uniformly. We report the absolute scores. In contrast,
FunnelAudit asks for the smallest factual-preserving contingency that makes
$c$ pivotal in the registered incident. It returns a replayable witness: the
contingency-only world and the world that additionally changes $c$.

The coalition scores are informative: their pooled Spearman correlations with
$\rho$ are .847--.965. They are nevertheless unreliable substitutes for an
incident judgment, especially for exclusions. For causal inclusions, a
maximum Shapley--Shubik score selects a responsible control in
96.01--100\% of incidents after uniform tie-breaking, and the Banzhaf range is
86.79--100\%. For causal exclusions, these ranges fall to 25.26--60.52\% and
73.20--93.32\%, even though exclusion correlations remain high (.860--.943).
Beauty C3 is the clearest example: exclusion Shapley--Shubik has correlation
.885 but TopResp. only 25.26\%. This mismatch is possible because a coalition
score aggregates marginal edges whose source worlds need not preserve the
registered factual outcome; such edges cannot certify responsibility for that
incident. RQ4 therefore does not show that coalition values are poor measures
of average credit. It shows that they cannot replace a minimum-contingency
judgment and its replayable witness.

\begin{table}[t]
  \centering
  \caption{Outcome-stratified agreement of Shapley--Shubik and Banzhaf scores
    with exact responsibility. Cells show inclusion/exclusion. Corr. is Spearman
    correlation with $\rho$ over all pairs in the stratum; TopResp. is the expected
    tie-broken probability that a maximum-score control is responsible in a causal
  incident.}
  \label{tab:coalition-new}
  \scriptsize
  \resizebox{\columnwidth}{!}{%
    \begin{tabular}{@{}llrrrr@{}}
      \toprule
      & & \multicolumn{2}{c}{Shapley--Shubik} & \multicolumn{2}{c}{Banzhaf} \\
      Dataset & Case & Corr. & TopResp. (\%) & Corr. & TopResp. (\%) \\
      \midrule
      \multirow{3}{*}{ML-1M}
      & C1 & 1.000/.899 & 100.00/48.62 & 1.000/.907 & 100.00/93.32 \\
      & C2 & .805/.865 & 99.53/60.52 & .565/.865 & 98.83/76.11 \\
      & C3 & .899/.860 & 99.20/46.52 & .453/.868 & 99.20/89.97 \\
      \midrule
      \multirow{3}{*}{Beauty}
      & C1 & 1.000/.931 & 100.00/30.39 & 1.000/.934 & 100.00/88.80 \\
      & C2 & .820/.900 & 100.00/45.03 & .714/.901 & 100.00/73.87 \\
      & C3 & .810/.885 & 99.51/25.26 & .685/.889 & 98.05/84.16 \\
      \midrule
      \multirow{3}{*}{Retail}
      & C1 & 1.000/.941 & 100.00/41.57 & 1.000/.943 & 100.00/92.68 \\
      & C2 & .794/.908 & 99.86/45.24 & .700/.908 & 99.40/80.90 \\
      & C3 & .810/.916 & 96.01/26.56 & .635/.917 & 86.79/73.20 \\
      \bottomrule
    \end{tabular}%
  }
\end{table}

\subsection{RQ5: Independent checks reproduce world construction and judgment}

RQ5 asks whether the reported judgments could be artifacts of the primary
compiler or extractor. We therefore test their two possible failure points
separately. First, an independently implemented literal replay starts from the
saved route outputs and model scores without importing the compiler. For 64
deterministically selected users per dataset, it reconstructs every eligible
target under all 512 C1 worlds, 1,024 C2 worlds, and 512 C3 worlds. All
9,121,792 target--world outcomes agree exactly with the compiled lattices.

Second, two checks start from those outcome lattices and independently test the
minimum-witness search. A per-pair exhaustive scan agrees with the primary
extractor on $\kappa$ and the canonical witness for 4,608 deterministically
sampled incident--control pairs. A generic selector MILP agrees on cause status
and, where a cause exists, minimum contingency size for 1,152 stratified pairs,
covering responsible and non-responsible inclusions and exclusions; every
feasible MILP witness also satisfies the pivotality conditions when checked
directly in the compiled lattice. Literal replay thus tests recommender-specific
world construction, while exhaustive scan and MILP test responsibility
extraction. Their joint agreement checks both parts without asking any
implementation to certify itself.

\section{Conclusion}

Auditing why an item appeared in, or was excluded from, a recommendation is not
the same as explaining its final ranking score. In a multi-route pipeline,
overlapping retrieval, allocation, fusion, and downstream ranking determine
whether a control becomes pivotal. \system{} makes this question executable
through an accountability contract that fixes the incident, accountable
controls, permitted reference actions, and replay semantics. It then applies
graded actual responsibility to the resulting counterfactual serving worlds
and returns a minimum-contingency judgment with a replayable witness. The causal
rule and optimization primitives are inherited; the contribution is the
recommender-specific specification and construction of the worlds and evidence
needed to use that rule in a serving audit.

Across 258,809 incidents from three real interaction datasets, most responsible
control--incident pairs are invisible to leave-one-control-out testing:
92.55--99.64\% require a nonempty contingency. Allocation and rank-aware fusion
also change responsible sets even when factual Top-$K$ outcomes rarely change.
Coalition contribution scores remain useful summaries, but their high aggregate
correlation with responsibility does not make them reliable substitutes for an
incident judgment, particularly for exclusions. Independent literal replay,
per-pair exhaustive search, and selector MILP reproduce the parts of the audit
they are designed to check.

Our evidence is limited to deterministic, binary route and allocator controls
in reconstructed two-stage funnels. RRF and SASRec are replayed but not audited,
and the results do not establish production prevalence, moral blame, legal
liability, or a remedy. Extending the contract to multi-valued quotas,
stochastic policies, downstream model controls, and larger registries remains
future work. Within the registered scope, the central requirement is clear: a
defensible accountability judgment needs both serving-faithful counterfactuals
and a checkable incident-level witness.

\bibliographystyle{ACM-Reference-Format}
\bibliography{references}

@article{chockler2004responsibility,
  author = {Hana Chockler and Joseph Y. Halpern},
  title = {Responsibility and Blame: A Structural-Model Approach},
  journal = {Journal of Artificial Intelligence Research},
  volume = {22},
  pages = {93--115},
  year = {2004},
  doi = {10.1613/jair.1391}
}

@article{bovens2007accountability,
  author = {Mark Bovens},
  title = {Analysing and Assessing Accountability: A Conceptual Framework},
  journal = {European Law Journal},
  volume = {13},
  number = {4},
  pages = {447--468},
  year = {2007},
  doi = {10.1111/j.1468-0386.2007.00378.x}
}

@article{diakopoulos2016accountability,
  author = {Nicholas Diakopoulos},
  title = {Accountability in Algorithmic Decision Making},
  journal = {Communications of the ACM},
  volume = {59},
  number = {2},
  pages = {56--62},
  year = {2016},
  doi = {10.1145/2844110}
}

@inproceedings{wieringa2020account,
  author = {Maranke Wieringa},
  title = {What to Account for When Accounting for Algorithms: A Systematic Literature Review on Algorithmic Accountability},
  booktitle = {Proceedings of the 2020 Conference on Fairness, Accountability, and Transparency},
  pages = {1--18},
  year = {2020},
  doi = {10.1145/3351095.3372833}
}

@inproceedings{raji2020closing,
  author = {Inioluwa Deborah Raji and Andrew Smart and Rebecca N. White and Margaret Mitchell and Timnit Gebru and Ben Hutchinson and Jamila Smith-Loud and Daniel Theron and Parker Barnes},
  title = {Closing the {AI} Accountability Gap: Defining an End-to-End Framework for Internal Algorithmic Auditing},
  booktitle = {Proceedings of the 2020 Conference on Fairness, Accountability, and Transparency},
  pages = {33--44},
  year = {2020},
  doi = {10.1145/3351095.3372873}
}

@inproceedings{cobbe2021reviewable,
  author = {Jennifer Cobbe and Michelle Seng Ah Lee and Jatinder Singh},
  title = {Reviewable Automated Decision-Making: A Framework for Accountable Algorithmic Systems},
  booktitle = {Proceedings of the 2021 ACM Conference on Fairness, Accountability, and Transparency},
  pages = {598--609},
  year = {2021},
  doi = {10.1145/3442188.3445921}
}

@inproceedings{kroll2021traceability,
  author = {Joshua A. Kroll},
  title = {Outlining Traceability: A Principle for Operationalizing Accountability in Computing Systems},
  booktitle = {Proceedings of the 2021 ACM Conference on Fairness, Accountability, and Transparency},
  pages = {758--771},
  year = {2021},
  doi = {10.1145/3442188.3445937}
}

@inproceedings{cooper2022algorithmic,
  author = {A. Feder Cooper and Emanuel Moss and Benjamin Laufer and Helen Nissenbaum},
  title = {Accountability in an Algorithmic Society: Relationality, Responsibility, and Robustness in Machine Learning},
  booktitle = {Proceedings of the 2022 ACM Conference on Fairness, Accountability, and Transparency},
  pages = {864--876},
  year = {2022},
  doi = {10.1145/3531146.3533150}
}

@article{meliou2010complexity,
  author = {Alexandra Meliou and Wolfgang Gatterbauer and Katherine F. Moore and Dan Suciu},
  title = {The Complexity of Causality and Responsibility for Query Answers and Non-Answers},
  journal = {Proceedings of the VLDB Endowment},
  volume = {4},
  number = {1},
  pages = {34--45},
  year = {2010},
  doi = {10.14778/1880172.1880176}
}

@inproceedings{meliou2010functional,
  author = {Alexandra Meliou and Wolfgang Gatterbauer and Katherine F. Moore and Dan Suciu},
  title = {Why So? or Why No? Functional Causality for Explaining Query Answers},
  booktitle = {Proceedings of the 4th International Workshop on Management of Uncertain Data},
  pages = {3--17},
  year = {2010},
  url = {https://arxiv.org/abs/0912.5340}
}

@inproceedings{verdeaux2020causality,
  author = {Will\`eme Verdeaux and Cl\'ement Moreau and Nicolas Labroche and Patrick Marcel},
  title = {Causality-Based Explanations in Multi-Stakeholder Recommendations},
  booktitle = {International Workshop on Explainability for Trustworthy ML Pipelines},
  series = {CEUR Workshop Proceedings},
  volume = {2578},
  year = {2020},
  url = {https://ceur-ws.org/Vol-2578/ETMLP2.pdf}
}

@article{bellogin2021accountability,
  author = {Alejandro Bellog\'{i}n and Alan Said},
  title = {Improving Accountability in Recommender Systems Research Through Reproducibility},
  journal = {User Modeling and User-Adapted Interaction},
  volume = {31},
  pages = {941--977},
  year = {2021},
  doi = {10.1007/s11257-021-09302-x}
}

@article{li2025acfr,
  author = {Jie Li and Yongli Ren and Mark Sanderson and Ke Deng},
  title = {Explaining Recommendation Fairness from a User/Item Perspective},
  journal = {ACM Transactions on Information Systems},
  volume = {43},
  number = {1},
  articleno = {17},
  numpages = {30},
  year = {2025},
  doi = {10.1145/3698877}
}

@article{messmer2023auditing,
  author = {Anna-Katharina Me{\ss}mer and Martin Degeling},
  title = {Auditing Recommender Systems: Putting the {DSA} into Practice with a Risk-Scenario-Based Approach},
  journal = {arXiv preprint arXiv:2302.04556},
  year = {2023},
  url = {https://arxiv.org/abs/2302.04556}
}

@article{sharma2024unified,
  author = {Vibhhu Sharma and Shantanu Gupta and Nil-Jana Akpinar and Zachary C. Lipton and Liu Leqi},
  title = {A Unified Causal Framework for Auditing Recommender Systems for Ethical Concerns},
  journal = {arXiv preprint arXiv:2409.13210},
  year = {2024},
  url = {https://arxiv.org/abs/2409.13210}
}

@article{singh2019decision,
  author = {Jatinder Singh and Jennifer Cobbe and Chris Norval},
  title = {Decision Provenance: Harnessing Data Flow for Accountable Systems},
  journal = {IEEE Access},
  volume = {7},
  pages = {6562--6574},
  year = {2019},
  doi = {10.1109/ACCESS.2018.2887201}
}

@inproceedings{huang2025multichannel,
  author = {Junjie Huang and Jiarui Qin and Jianghao Lin and Ziming Feng and Weinan Zhang and Yong Yu},
  title = {Unleashing the Potential of Multi-Channel Fusion in Retrieval for Personalized Recommendations},
  booktitle = {Proceedings of the ACM Web Conference 2025},
  pages = {483--494},
  year = {2025},
  doi = {10.1145/3696410.3714753}
}

@inproceedings{covington2016youtube,
  author = {Paul Covington and Jay Adams and Emre Sargin},
  title = {Deep Neural Networks for YouTube Recommendations},
  booktitle = {Proceedings of the 10th ACM Conference on Recommender Systems},
  pages = {191--198},
  year = {2016},
  doi = {10.1145/2959100.2959190}
}

@inproceedings{wang2020cold,
  author = {Zhe Wang and Liqin Zhao and Biye Jiang and Guorui Zhou and Xiaoqiang Zhu and Kun Gai},
  title = {{COLD}: Towards the Next Generation of Pre-Ranking System},
  booktitle = {Proceedings of the 2nd Workshop on Deep Learning Practice for High-Dimensional Sparse Data},
  year = {2020},
  url = {https://arxiv.org/abs/2007.16122}
}

@inproceedings{sarwar2001itembased,
  author = {Badrul Sarwar and George Karypis and Joseph Konstan and John Riedl},
  title = {Item-Based Collaborative Filtering Recommendation Algorithms},
  booktitle = {Proceedings of the 10th International Conference on World Wide Web},
  pages = {285--295},
  year = {2001},
  doi = {10.1145/371920.372071}
}

@inproceedings{resnick1994grouplens,
  author = {Paul Resnick and Neophytos Iacovou and Mitesh Suchak and Peter Bergstrom and John Riedl},
  title = {{GroupLens}: An Open Architecture for Collaborative Filtering of Netnews},
  booktitle = {Proceedings of the 1994 ACM Conference on Computer Supported Cooperative Work},
  pages = {175--186},
  year = {1994},
  doi = {10.1145/192844.192905}
}

@inproceedings{cormack2009rrf,
  author = {Gordon V. Cormack and Charles L. A. Clarke and Stefan B{\"u}ttcher},
  title = {Reciprocal Rank Fusion Outperforms Condorcet and Individual Rank Learning Methods},
  booktitle = {Proceedings of the 32nd International ACM SIGIR Conference on Research and Development in Information Retrieval},
  pages = {758--759},
  year = {2009},
  doi = {10.1145/1571941.1572114}
}

@inproceedings{lundberg2017shap,
  author = {Scott M. Lundberg and Su-In Lee},
  title = {A Unified Approach to Interpreting Model Predictions},
  booktitle = {Advances in Neural Information Processing Systems},
  volume = {30},
  year = {2017},
  url = {https://proceedings.neurips.cc/paper/2017/hash/8a20a8621978632d76c43dfd28b67767-Abstract.html}
}

@inproceedings{wang2023banzhaf,
  author = {Jiachen T. Wang and Ruoxi Jia},
  title = {Data Banzhaf: A Robust Data Valuation Framework for Machine Learning},
  booktitle = {Proceedings of the 26th International Conference on Artificial Intelligence and Statistics},
  series = {Proceedings of Machine Learning Research},
  volume = {206},
  pages = {6388--6421},
  year = {2023},
  url = {https://proceedings.mlr.press/v206/wang23e.html}
}

@article{yao2022counterfactual,
  author = {Yuanshun Yao and Chong Wang and Hang Li},
  title = {Learning to Counterfactually Explain Recommendations},
  journal = {arXiv preprint arXiv:2211.09752},
  year = {2022},
  url = {https://arxiv.org/abs/2211.09752}
}

@inproceedings{tan2021counter,
  author = {Juntao Tan and Shuyuan Xu and Yingqiang Ge and Yunqi Li and Xu Chen and Yongfeng Zhang},
  title = {Counterfactual Explainable Recommendation},
  booktitle = {Proceedings of the 30th ACM International Conference on Information and Knowledge Management},
  pages = {1784--1793},
  year = {2021},
  doi = {10.1145/3459637.3482420}
}

@inproceedings{ghazimatin2020prince,
  author = {Azin Ghazimatin and Oana Balalau and Rishiraj Saha Roy and Gerhard Weikum},
  title = {{PRINCE}: Provider-Side Interpretability with Counterfactual Explanations in Recommender Systems},
  booktitle = {Proceedings of the 13th International Conference on Web Search and Data Mining},
  pages = {196--204},
  year = {2020},
  doi = {10.1145/3336191.3371824}
}

@inproceedings{stratigi2020whynot,
  author = {Maria Stratigi and Katerina Tzompanaki and Kostas Stefanidis},
  title = {Why-Not Questions and Explanations for Collaborative Filtering},
  booktitle = {Web Information Systems Engineering},
  series = {Lecture Notes in Computer Science},
  volume = {12343},
  pages = {301--315},
  year = {2020},
  doi = {10.1007/978-3-030-62008-0_21}
}

@inproceedings{attolou2024whynot,
  author = {Herv{\'e} Madelein Attolou and Katerina Tzompanaki and Kostas Stefanidis and Dimitris Kotzinos},
  title = {Why-Not Explainable Graph Recommender},
  booktitle = {Proceedings of the 40th IEEE International Conference on Data Engineering},
  pages = {2245--2257},
  year = {2024},
  doi = {10.1109/ICDE60146.2024.00178}
}

@inproceedings{mohammadi2025beyond,
  author = {Amir Reza Mohammadi and Andreas Peintner and Michael M{\"u}ller and Eva Zangerle},
  title = {Beyond Top-1: Addressing Inconsistencies in Evaluating Counterfactual Explanations for Recommender Systems},
  booktitle = {Proceedings of the 19th ACM Conference on Recommender Systems},
  pages = {515--520},
  year = {2025},
  doi = {10.1145/3705328.3748028}
}

@inproceedings{kim2026trace,
  author = {Ungsik Kim and Sang-Min Choi and Gun-Woo Kim and Suwon Lee},
  title = {{TRACE}: Targeted Ranking-Aware Counterfactual Explanation for Sequential Recommendation},
  booktitle = {Proceedings of the 20th ACM Conference on Recommender Systems},
  year = {2026},
  doi = {10.1145/3773078.3831796}
}

@inproceedings{kang2018sasrec,
  author = {Wang-Cheng Kang and Julian McAuley},
  title = {Self-Attentive Sequential Recommendation},
  booktitle = {2018 IEEE International Conference on Data Mining},
  pages = {197--206},
  year = {2018},
  doi = {10.1109/ICDM.2018.00035}
}

@inproceedings{rendle2009bpr,
  author = {Steffen Rendle and Christoph Freudenthaler and Zeno Gantner and Lars Schmidt-Thieme},
  title = {{BPR}: Bayesian Personalized Ranking from Implicit Feedback},
  booktitle = {Proceedings of the 25th Conference on Uncertainty in Artificial Intelligence},
  pages = {452--461},
  year = {2009},
  url = {https://arxiv.org/abs/1205.2618}
}

@inproceedings{he2017neumf,
  author = {Xiangnan He and Lizi Liao and Hanwang Zhang and Liqiang Nie and Xia Hu and Tat-Seng Chua},
  title = {Neural Collaborative Filtering},
  booktitle = {Proceedings of the 26th International Conference on World Wide Web},
  pages = {173--182},
  year = {2017},
  doi = {10.1145/3038912.3052569}
}

@inproceedings{he2020lightgcn,
  author = {Xiangnan He and Kuan Deng and Xiang Wang and Yan Li and Yongdong Zhang and Meng Wang},
  title = {{LightGCN}: Simplifying and Powering Graph Convolution Network for Recommendation},
  booktitle = {Proceedings of the 43rd International ACM SIGIR Conference on Research and Development in Information Retrieval},
  pages = {639--648},
  year = {2020},
  doi = {10.1145/3397271.3401063}
}

@inproceedings{mao2021simplex,
  author = {Kelong Mao and Jieming Zhu and Jinpeng Wang and Quanyu Dai and Zhenhua Dong and Xi Xiao and Xiuqiang He},
  title = {{SimpleX}: A Simple and Strong Baseline for Collaborative Filtering},
  booktitle = {Proceedings of the 30th ACM International Conference on Information and Knowledge Management},
  pages = {1243--1252},
  year = {2021},
  doi = {10.1145/3459637.3482297}
}

@article{harper2015movielens,
  author = {F. Maxwell Harper and Joseph A. Konstan},
  title = {The MovieLens Datasets: History and Context},
  journal = {ACM Transactions on Interactive Intelligent Systems},
  volume = {5},
  number = {4},
  pages = {19:1--19:19},
  year = {2015},
  doi = {10.1145/2827872}
}

@inproceedings{mcauley2015amazon,
  author = {Julian McAuley and Christopher Targett and Qinfeng Shi and Anton van den Hengel},
  title = {Image-Based Recommendations on Styles and Substitutes},
  booktitle = {Proceedings of the 38th International ACM SIGIR Conference on Research and Development in Information Retrieval},
  pages = {43--52},
  year = {2015},
  doi = {10.1145/2766462.2767755}
}

@misc{chen2012onlineretail,
  author = {Daqing Chen},
  title = {Online Retail II},
  year = {2012},
  publisher = {UCI Machine Learning Repository},
  doi = {10.24432/C5CG6D}
}

@book{nemhauser1988integer,
  author = {George L. Nemhauser and Laurence A. Wolsey},
  title = {Integer and Combinatorial Optimization},
  publisher = {John Wiley \& Sons},
  year = {1988},
  doi = {10.1002/9781118627372},
  isbn = {9780471828198}
}

@article{rubinstein1999cem,
  author = {Reuven Y. Rubinstein},
  title = {The Cross-Entropy Method for Combinatorial and Continuous Optimization},
  journal = {Methodology and Computing in Applied Probability},
  volume = {1},
  pages = {127--190},
  year = {1999},
  doi = {10.1023/A:1010091220143}
}

@inproceedings{waiwitlikhit2024trustless,
  author = {Suppakit Waiwitlikhit and Ion Stoica and Yi Sun and Tatsunori Hashimoto and Daniel Kang},
  title = {Trustless Audits without Revealing Data or Models},
  booktitle = {Proceedings of the 41st International Conference on Machine Learning},
  series = {Proceedings of Machine Learning Research},
  volume = {235},
  pages = {49808--49821},
  year = {2024},
  url = {https://proceedings.mlr.press/v235/waiwitlikhit24a.html}
}

@article{zhu2026forget,
  author = {Zhihao Zhu and Yi Yang and Yangyang Fan and Defu Lian},
  title = {Forget Me If You Can: Auditing User Data Revocation in Recommendation Systems},
  journal = {Information Systems Research},
  year = {2026},
  note = {Published online},
  doi = {10.1287/isre.2024.1179}
}

@inproceedings{hsu2025models,
  author = {Brian Hsu and Cyrus DiCiccio and Natesh S. Pillai and Hongseok Namkoong},
  title = {From Models to Systems: A Comprehensive Framework for {AI} System Fairness in Compositional Recommender Systems},
  booktitle = {Proceedings of the Algorithmic Fairness Through the Lens of Metrics and Evaluation},
  series = {Proceedings of Machine Learning Research},
  volume = {279},
  pages = {8--37},
  year = {2025},
  url = {https://proceedings.mlr.press/v279/hsu25a.html}
}

@misc{linkedin2024semantic,
  author = {Xin Yang and Rachel Zheng and Madhumitha Mohan and Sonali Bhadra and Pansul Bhatt and Lingyu Zhang and Rupesh Gupta},
  title = {Introducing Semantic Capability in {LinkedIn}'s Content Search Engine},
  howpublished = {LinkedIn Engineering Blog},
  year = {2024},
  month = aug,
  url = {https://www.linkedin.com/blog/engineering/search/introducing-semantic-capability-in-linkedins-content-search-engine},
  note = {Accessed 2026-08-27}
}

\end{document}